\documentclass[reprint,aps,prl,twocolumn]{revtex4-2}
\usepackage[british,UKenglish]{babel}
\usepackage{amsmath,amssymb,amsopn,bbm}
\usepackage{graphicx}
\usepackage{epstopdf}
\usepackage{hyperref}
\hypersetup{
    colorlinks=true,
    citecolor=cyan,
    linkcolor=red,
    filecolor=orange,      
    urlcolor=gray,
    pdfpagemode=FullScreen,
    }
\usepackage[usenames,dvipsnames]{xcolor}
\begin{document}
%
%
\title{Quantum Machine Learning With Canonical Variables}
\author{J. Fuentes}
\email[Correspondence:~]{jesus.fuentes [at] uni.lu}
\affiliation{LCSB, University of Luxembourg \\ 6, Avenue Du Swing, \\
L-4364 Belval, Luxembourg}
%
%
\begin{abstract}


Utilising dynamic electromagnetic field control over charged particles serves as the basis for a quantum machine learning platform that operates on observables rather than directly on states.
Such a platform can be physically realised in ion traps or particle confinement devices that utilise electromagnetic fields as the source of control.
The electromagnetic field acts as the ansatz within the learning algorithm.
The models discussed are exactly solvable, with exact solutions serving as precursors for learning tasks to emerge, including regression and classification algorithms as particular cases.
This approach is considered in terms of canonical variables with semi-classical behaviour, disregarding relativistic degrees of freedom.

\end{abstract}
%
%
%
%
\maketitle
%
%
\section{Introduction}

A machine learning algorithm consists of fitting the parameters of a sufficiently smooth ansatz through an optimisation process to approximate the pattern in the data, if any, identifying how $x$ maps to $y$.
Therefore, if the algorithm has accurately learned (or fitted) the map $f(x) = y$, then any unseen data of the same kind could be predicted.

Any learning method comprises three fundamental components: the data, the ansatz or model, and the optimisation scheme.
The type of data determines whether the learning algorithm should be modelled by a classical or quantum ansatz, while the optimisation can be either classical or quantum, irrespective of the data’s nature. 
In principle, classical data can be mapped to a quantum ansatz, but not all quantum data can be mapped to a classical ansatz.

For classical learning algorithms, the ansatz is intangible, being it modelled as a smooth function, such as a polynomial or a neural network.
The entire algorithm is instantiated as code, and then is transformed into a low-level language via a compiler or interpreter.
At that level, the instructions are executed by the central processing unit, where the arithmetic operations and logic decisions required by the optimisation protocol are processed as 1s and 0s by arrays of transistors.

In contrast, the ansatz in quantum learning algorithms is inherently physical, modelled through quantum circuits.
Initially instantiated as classical digital code, the algorithm is transformed into quantum instructions via a quantum compiler, enabling qubit-level processing.
Classical data is encoded into quantum states for this process and then decoded back to classical form for interpretation.
In cases where the data is quantum, it remains in quantum states, allowing direct manipulation by quantum algorithms.
Regardless the type of data---quantum or classical---decoding qubits into the digital domain is necessary for interpreting the results in the classical landscape.

Optimisation in this context can be carried out using either quantum or classical algorithms.
Quantum optimisation is usually performed using techniques like the quantum approximate optimisation algorithm \cite{farhi14} or the variational quantum eigensolver \cite{peruzzo2014} to find optimal parameters.
Classical optimisation schemes, such as gradient descent or Bayesian optimisation \cite{mockus94}, can also be used, and often a hybrid approach is employed \cite{benedetti19, mcclean16} where classical and quantum computations complement each other.

The field of quantum machine learning has seen rapid advancements \cite{rebentrost14,dunjko16,mitarai18}.
The implementation of quantum algorithms for supervised learning tasks has demonstrated the potential for significant improvements over classical methods \cite{schuld18}.
Research also indicates that quantum-enhanced clustering and quantum neural networks can achieve more accurate data segmentation and faster learning rates compared to their classical counterparts \cite{biamonte17}. Furthermore, experimental demonstrations, such as the implementation of quantum neural networks on actual quantum processors, reflect the feasibility and ongoing progress in the field \cite{havlicek19}.

Nevertheless, another form of quantum machine learning is possible that does not involve qubit manipulation for learning tasks to be performed.
Efforts in this direction have been conceptualised as quantum machine learning with continuous variables \cite{braunstein05,killoran19}.
However, little attention has been given to the dynamical operations that generate learning protocols, and even less to the exactly solvable quantum systems that permit such learning tasks to be realised.
These are the primary concerns that are addressed in this letter.

Below, the author presents learning algorithms that can be embedded in devices dedicated to ion confinement.
In this form of learning algorithms, attention is given to operators driven directly by time-dependent elastic potentials, resulting in algorithms tailored for quantum time-dependent problems.
Hence, in this scenario, the ion trap serves as the processing framework while the elastic potential serves as the ansatz. 
This approach is discussed in terms of canonical variables with semi-classical behaviour, disregarding relativistic degrees of freedom.
Still, as demonstrated in the appendix, the applicability of this scheme extends beyond canonical observables and includes the time evolution of a Gaussian wave packet.
Two types of supervised learning are investigated: regression and classification.
The quantum model that allows these operations to occur is exactly solvable, and its exact solutions favour the development of such operations.
Accompanying this letter are Python code and synthetic datasets to test and reproduce the numerical algorithms discussed.

In what follows, the utilisation of dimensionless variables will be particularly convenient for simplifying the analysis.
This can be done through $t = \omega' t’$, $p = \sqrt{1/m\omega\hbar} p’$, and $x = \sqrt{m\omega/\hbar} x’$, where $t’$ is the time in seconds, $p’$ is the momentum operator for an ion with charge $e$ and mass $m$ moving along the coordinate $x’$, $\hbar$ is a constant with units of action, and $\omega' = 1/\tau'$, with $\tau'$ being the operational time scale in seconds.
Upon these substitutions the canonical quantisation $[x, p] = \mathrm{i}$ is met.

%
%
\section{The motion of a confined ion}

Charged particles can be confined within regions dominated by time-varying electric or magnetic fields, or a combination thereof.
In the domain of electric trapping mechanisms, Paul's theory of ion confinement \cite{paul90} discusses how particles injected into devices with hyperbolic shapes are elastically focused along specific coordinates, $\mathbf{x}' = (x' ~ y' ~ z')$. Within a typical Paul trap with radius $r_0$, the influence of retarded potentials on the ion is usually negligible, permitting the approximation of quadrupole electric potentials in the form of $\Phi = \phi(t')/2 r_0^2 \cdot \mathbf{x}'\Omega \mathbf{x}'^\intercal$.
Here, $\phi(t')$ denotes an elastic scalar field produced by modulated voltages, and $\Omega = \operatorname{diag}(\alpha, \beta, \gamma)$ is a real, traceless matrix that characterises the geometry of the trap.

These potentials, being quadratic forms, allow the motion of the ion to be analysed independently for each coordinate.
In this setting, using the aforementioned substitutions for dimensionless variables, the motion of an ion along $Ox$ can be described by Hamiltonians of the form:
\begin{equation}\label{hamiltonian}
H_\beta = \frac{H}{\hbar\omega} = \frac{1}{2}p^2 + \frac{\beta(t)}{2}x^2,
\end{equation}
where $\beta(t) = e \phi(\omega t) /m \omega^2 r_0^2$ represents the dimensionless elastic potential.


Quadratic Hamiltonians are not unique to ions moving in quadrupole traps; they also arise when charged particles move in a time-varying magnetic fields.
In such fields, ions spiral around field lines due to the Lorentz force, which stabilises their motion perpendicular to the magnetic field direction.
Consider a time-dependent magnetic induction field $\mathbf{B}' = B(t') \hat{\mathbf{z}}$ in a symmetric gauge $\mathbf{A}' = (\mathbf{B}' \times \mathbf{x}') / 2$, which reflects a cylindrical geometry akin to the topology of $\mathbb{R}\times S^1$ generated by $\Omega = \operatorname{diag}(1,1,-2)$. 

In the absence of other forces and spin, the Peierls substitution $[\mathbf{p}' - e \mathbf{A}'/c]^2/2m$ directly gives rise to a Hamiltonian that unfolds into three components:
The parallel component, $H_\parallel'$, describes the motion of a free particle along the $Oz$ direction;
the rotational component, $H_\circlearrowleft'$, generates symplectic rotations around $Oz$, signifying that angular momentum in this direction is an integral of the motion;
and the oscillatory component, $H_\perp'$, which governs oscillations in the $xOy$ plane.
Hence, the dimensionless Hamiltonian $H_\perp/\hbar\omega$ takes the form \eqref{hamiltonian} for each coordinate, independent of $B(t')$ oscillates periodically with frequency $\omega$ or not.
In this instance, the elastic amplitude is defined by $\beta(t) = (\hbar / 2m\omega\ell_B^2(\omega t))^2$, where $\ell_B(\omega t)$ is the time-dependent magnetic length.

Therefore, in both electric and magnetic scenarios, the profile $\beta(t)$ should dictate that the sequence of field applications either captures or expels ions, necessitating a chain of focusing and defocusing operations for the particle to remain confined during the operation's time interval $\tau$, as outlined in \cite{mielnik86}.
Since a bound particle typically resides near an equilibrium point, it encounters restoring forces proportional to its displacement from this point.
Consequently, the focusing and defocusing mechanisms depend on the field’s strength along a particular coordinate, counterbalancing the free motion with oscillatory and rotational effects.
As will be demonstrated, the creation of these sequences, derived from the exact solutions to the evolution problem governed by $H_\beta$, will allow the generation of learning operations in an analogue form of quantum machine learning.

%
%
\section{The Exact Solutions}

Resolving the evolution problem defined by \eqref{hamiltonian} is relatively simple.
For a well-behaved $\beta(t)$ over the time interval $[t_0,t]$, the Heisenberg equations of motion simplify to $\mathrm{d}p/\mathrm{d}t = -\beta(t) x$ and $\mathrm{d}x/\mathrm{d}t = p$, which share the same structural form in both classical and quantum regimes, as the $H_\beta$ is quadratic.
The solutions to this set of equations can be represented as a linear transformation of the canonical variables, $Q=(x ~ p)^\intercal$, promoted by evolution matrices $u(t,t_0) \in \operatorname{Sp}(2,\mathbb{R})$.
Namely, the ion’s canonical observables evolve according to $Q = u(t,t_0) Q_0$, with $Q_0 = (x_0 ~ p_0)^\intercal = (x(t_0) ~  p(t_0))^\intercal$. 
Whereas the evolution matrix $u(t,t_0)$ is determined by integrating the Cauchy problem:
\begin{equation}\label{cauchy}
\frac{\mathrm{d}u(t,t_0)}{\mathrm{d}t} = \Lambda(t)u(t,t_0), \quad u(t_0,t_0) = \mathbbm{1},
\end{equation}
where $\Lambda(t) = \operatorname{antidiag}(1, -\beta(t))$.

The solution to \eqref{cauchy} defines three types of ion motion.
By denoting $\Gamma(t) = \operatorname{Tr} u(t,t_0)$, and because $u(t,t_0)$ is symplectic, the eigenvalues of the evolution matrix are $\kappa = (\Gamma(t) \pm \sqrt{\Gamma^2(t) - 4})/2$.
These eigenvalues correspond to a unique type of motion: for $\vert \Gamma(t) \vert > 2$, the eigenvalues $\kappa$ are real, causing the ion to defocus \cite{fuentes23}; for $\vert \Gamma(t) \vert = 2$, an edge (separatrix) region is defined and, although $\kappa$ is also real, stable motion may occur \cite{mielnik14}; for $\vert \Gamma(t) \vert < 2$, $\kappa$ is complex, causing the ion to oscillate and focus around a coordinate or axis \cite{mielnik10,mielnik11,fuentes20}.

The time evolution problem in \eqref{cauchy} admits exact solutions through algebraic arguments, without the need for any adiabatic invariant scheme \cite{mielnik13,mielnik14}.
Two cases are readily integrable: when the elastic field $\beta(t)$ is even over the interval $t \in [-\tau, \tau]$, and when $\beta(t)$ is odd over the same interval \cite{fuentes23}.
However, the latter case will not be considered here; all the learning operations discussed will be derived assuming $\beta(t)$ is an even function.

If $\beta(-t) = \beta(t)$ for $t \in [-\tau, \tau]$, then $H_\beta(-t) = H_\beta(t)$. This scenario can occur for either electric or magnetic fields, the latter iff $\beta(t) \geq 0$. Therefore, the Cauchy problem \eqref{cauchy} becomes:
\begin{equation*}
\frac{\mathrm{d}u(t,-\tau)}{\mathrm{d}t} = \frac{1}{2} \{\Lambda(t), u(t,-\tau)\}, \quad u(-\tau,-\tau) = \mathbbm{1}.
\end{equation*}
The exact solution to this system of linear evolution equations is given by \cite{mielnik13,fuentes23}:
\begin{equation}\label{solution}
\renewcommand{\arraystretch}{1.5}
u(t,-\tau) = \begin{pmatrix} \dot{\theta}_\mathcal{P}(t) & \theta_\mathcal{P}(t) \\ \frac{\dot{\theta}^2_\mathcal{P}(t)-1}{2\theta_\mathcal{P}(t)} & \dot{\theta}_\mathcal{P}(t) \end{pmatrix}.
\end{equation}

The function $\theta_\mathcal{P}(t)\in\mathbb{R}$ is an almost arbitrary function that must be non-singular and bounded in the interval $[-\tau,\tau]$, with $\mathcal{P}$ denoting a set of real parameters.
Such behaviour is assured if, when $\theta_\mathcal{P}(t) = 0$, then $\dot{\theta}_\mathcal{P}(t) = \pm 1$; and if, when both $\beta(t) \neq 0$ and $\theta_\mathcal{P}(t) \neq 0$ while their time derivatives are simultaneously zero, then $\dddot{\theta}_\mathcal{P}(t) = 0$. 
The first condition implies $\Gamma(t) = \pm 2$, which indicates critical points where the system's stability can switch, representing thresholds in the dynamic response of the ion.
In the second condition, the absence of second-order time derivatives of $\Gamma(t)$, when both $\beta(t)$ and $\theta_\mathcal{P}(t)$ are non-zero but not changing, indicates a steady state, thus suggesting a momentary equilibrium in the motion.

The elastic field is related to the $\theta_\mathcal{P}(t)$ function through a second-order, non-linear differential equation:
\begin{equation}\label{oscillator}
\ddot{\theta}_\mathcal{P}(t) + \frac{\beta(t)}{2}\theta_\mathcal{P}(t) = \frac{\dot{\theta}_\mathcal{P}^2(t)-1}{2\theta_\mathcal{P}(t)},
\end{equation}
where the right-hand side is interpreted as an effective self-driving component.
For instance, for a constant oscillatory field, $\beta(t)=\omega_0^2$, the solution is customary: $\theta_\mathcal{P}(t) = \theta_{\omega_0}(t) = \sin(\omega_0 t)/\omega_0$, leading \eqref{solution} to become a matrix of symplectic rotations with a period of $2\pi/\omega_0$.

Therefore, either $\beta(t)$ or $\theta_\mathcal{P}(t)$ can be chosen to produce exact solutions to \eqref{cauchy} in a kind of inverse problem. 
It is precisely the optimisation of $\theta_\mathcal{P}(t)$ that allows the development of quantum learning operations to drive the canonical variables, permitting classification or regression tasks.

%
%
\section{Ion Learning Algorithms}

In what follows, the notation $\hat{\Xi}$ and $\Xi$ will be used to distinguish between the output (target) and trained variables, in that order.
Specifically, $\{t, \hat{\Xi}\}$ represents a dataset containing information of a particle moving under the influence of an elastic potential $\beta(t)$ in an ion trap, where $t$ is the feature and $\hat{\Xi}$ is the target outcome.
The data in $\hat{\Xi}$ could be empirical, obtained from experimental observations, or synthetic, designed for a specific operation. 
In practice, if the target observable $\hat{\Xi}$ comes from an experiment, it is expected to contain noisy measurements.

Exact solutions to \eqref{cauchy} enable the development of supervised quantum learning algorithms.
In particular, two types of algorithms will be discussed in order: regression and classification.
In regression, continuous outcomes are predicted based on the motion described by the target variables $\hat{\Xi}$, representing a system’s set of quantum observables.
In contrast, classification focuses on categorising $\hat{\Xi}$ into discrete classes on the basis of predefined criteria, thus segmenting the canonical variables into distinct quantum classes.

\subsection{Regression Algorithms}

\begin{figure*}[t]
\centering
\includegraphics[width=\textwidth]{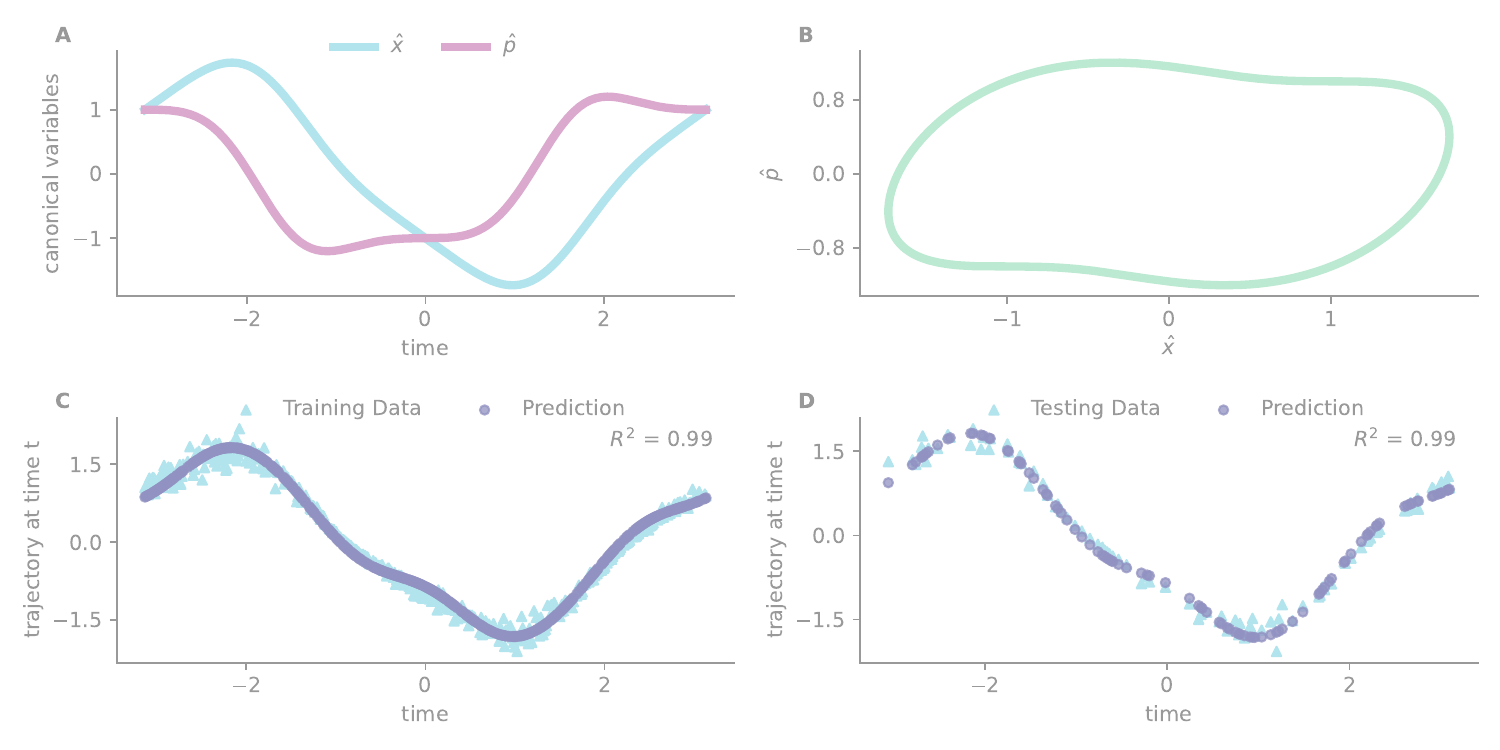}
\caption{Canonical variables utilised for training the regression model. Panel A: The noiseless position and momentum, $\hat{x}$ and $\hat{p}$, evolved over the time interval $[-\pi,\pi]$ by applying the evolution matrix derived from \eqref{cauchy} to the initial condition $(\hat{x}_0,\hat{p}_0) = (1, 1)$. These variables serve as synthetic data by adding $10\%$ noise. Panel B: Phase portrait of noiseless $\hat{x}$ and $\hat{p}$ showing an evolution loop. Panel C: The regression algorithm’s performance using noisy $\hat{x}$ as the target variable, evaluated on the training data, with an RMSE of $0.143$ and an $R^2$ of $0.99$. Panel D: Analogous evaluation on the testing set achieving an RMSE of $0.14$ and $R^2$ of $0.99$.}
\label{fig:regression}
\end{figure*}

The following regression algorithm trains an observable $\Xi$ to learn $\hat{\Xi}$ by iteratively minimising a cost function $\mathcal{C}_R$, which quantifies the discrepancy between them.
This optimisation process, classical or quantum, adjusts the parameters in $\theta_\mathcal{P}(t)$, to approximate $\Xi$ as accurately as possible towards $\hat{\Xi}$.

Such a learning algorithm can be formulated as $\min_{\Xi,\theta_\mathcal{P}(t)} \mathcal{C}_R$, subject to earlier discussed conditions on $\theta_\mathcal{P}(t)$ to ensure $\beta(t)$ is physically admissible.
The cost function used for this algorithm is defined as the root mean square error (RMSE), $\mathcal{C}_R = \sum_{j=1}^n (\hat{\Xi}_j - \Xi_j)^2/n$, where $n$ represents the number of time points in the target variable.

If, for instance, the target is the ion's trajectory, $\hat{\Xi} = \hat{x}$, the cost function becomes $\mathcal{C}_R = \sum_{j=1}^n (\hat{x}(t_j) - x(t_j))^2/n$, with $x(t_j) = \hat{x}_0\dot{\theta}_\mathcal{P}(t_j) + \hat{p}_0\theta_\mathcal{P}(t_j)$, resulting from the linear transformation $Q=u(t,-\tau)Q_0$ in accordance with \eqref{solution}, cf. \footnote{It is assumed that the initial conditions $\hat{Q}_0 = Q_0$.}.
Moreover, even without any previous knowledge of the momentum $\hat{p}$, this canonical observable can be predicted provided the parameters $\mathcal{P}$ have been optimised for $\hat{x}$.
The reason is that $\hat{x}$ and $\hat{p}$ are conjugate variables and both evolve in terms of $\theta_\mathcal{P}(t)$ according to \eqref{solution}.
Therefore, once one variable is learned, its counterpart can be predicted.

As the model is iteratively refined and updated through an optimisation process, the parameter space $\mathcal{P}$ is navigated  to identify the optimal parameters that minimise $\mathcal{C}_R$.
All the learning algorithms discussed in this letter will minimise the cost function by scanning the parameter space $\mathcal{P}$ using Bayesian classical optimisation \cite{mockus94}.
Further technical details about the algorithms' implementation can be found in the accompanying code \footnote{The code and databases that accompany this work can be freely accessed at \href{https://github.com/fuentesigma/ionlearning}{github.com/fuentesigma/ionlearning}}.

To illustrate how this works, synthetic data must be generated first. 
To this aim, the elastic field will be defined as the superposition of $M$ harmonic waves, namely $\beta(t) = \sum_{j=0}^M \beta_j \cos(\omega_j t)$, where $\beta_j$ are constant coefficients and $t \in [-2\pi, 2\pi]$.
For this numerical example, the first four terms in the sum suffice, with the amplitudes $\beta_0 = 25/24$, $\beta_1 = 1/11$, $\beta_2 = 37/36$, $\beta_3 = 1/12$, and the frequencies $\omega_0 = 0$, $\omega_1 = 4/25$, $\omega_2 = 2$, $\omega_3 = 4$.

Integrating the system of equations \eqref{cauchy} to account for this $\beta(t)$ yields an evolution matrix with complex eigenvalues, thus promoting focused motion. 
Applying this evolution matrix to the arbitrary initial condition $(\hat{x}_0, \hat{p}_0) = (1, 1)$, produces the time evolution of the coordinates and momentum, denoted as $\hat{x}$ and $\hat{p}$ over the interval $[-2\pi, 2\pi]$.
The evolved observables $\hat{x}$ and $\hat{p}$ exhibit periodic behaviour and manifest evolution loops over the time interval, meaning both return to their initial conditions at the end of the operation.
These variables are illustrated in panels A and B of Figure \ref{fig:regression}, computed using 500 uniformly distributed time points.
However, to evaluate the model’s ability to generalise, Gaussian noise at $10\%$ has been added to both observables.
Consequently, the (noisy) position $\hat{x}$ and momentum $\hat{p}$ can serve as synthetic data for training the regression model, with the dataset given by $\{t,\hat{x}\}$, in this instance.

Resolving a problem using machine learning, whether quantum or classical, requires the algorithm to have an ansatz as a starting approximation that can evolve to match the target solution.
In this case, the ansatz is given by $
\theta_\mathcal{P}(t)) = \sum_{j=1}^{N_\mathcal{P}} a_{2j-1} \sin([2j-1] t),$
where the parameter space is spanned by $\mathcal{P} = \{a_{2j-1} \, \vert \, j = 1, \ldots, N_\mathcal{P}\}$. 
Here, $N_\mathcal{P}$ denotes the number of coefficients, which can be either fixed arbitrarily or optimised through fine-tuning.
For this numerical experiment, $N_\mathcal{P}$ was set to 3, which is sufficient to approximate the behaviour of $\hat{x}$.

To initiate the training process, the dataset $\{t, \hat{x}\}$ is divided into training and testing subsets in an 80:20 ratio.
The partitioning is performed such that the samples are accurate representations of the overall dataset, maintaining the statistical properties and distributions of the data.
Subsequently, the training set is fed into the regression model, and the optimisation process begins.

To optimise the parameters $\mathcal{P}$, the fitting is performed on the training set, adjusting $\mathcal{P}$ to minimise the loss function.
Overfitting is avoided by stopping the optimisation automatically when the loss function shows no significant improvement.
This prevents the algorithm from merely memorising the training set and ensures the model generalises well to unseen data.
After learning the parameters $\mathcal{P}$, the model’s performance is evaluated on the testing subset, which consists of previously unseen data.
Panels C and D of Figure \ref{fig:regression} summarise the results of the model’s performance on both the training and testing subsets after 78 iterations.
As shown, the $R^2$ value is close to unity, indicating that the model explains almost all the variance in the data.

The exact solutions to \eqref{cauchy} also allow for the evolution of other quantities, such as the probability density in terms of a Gaussian wave packet \cite{fuentes23}.
Hence, this learning scheme is not limited to regression or classification using canonical variables. 
See the Supplemental Materials, where regression analysis is applied to the time evolution of a Gaussian wave packet under the influence of the same elastic potential $\beta(t)$ studied here, and where datasets include spatio-temporal features.

\subsection{Classification Algorithms}

\begin{figure*}[t]
\centering
\includegraphics[width=\textwidth]{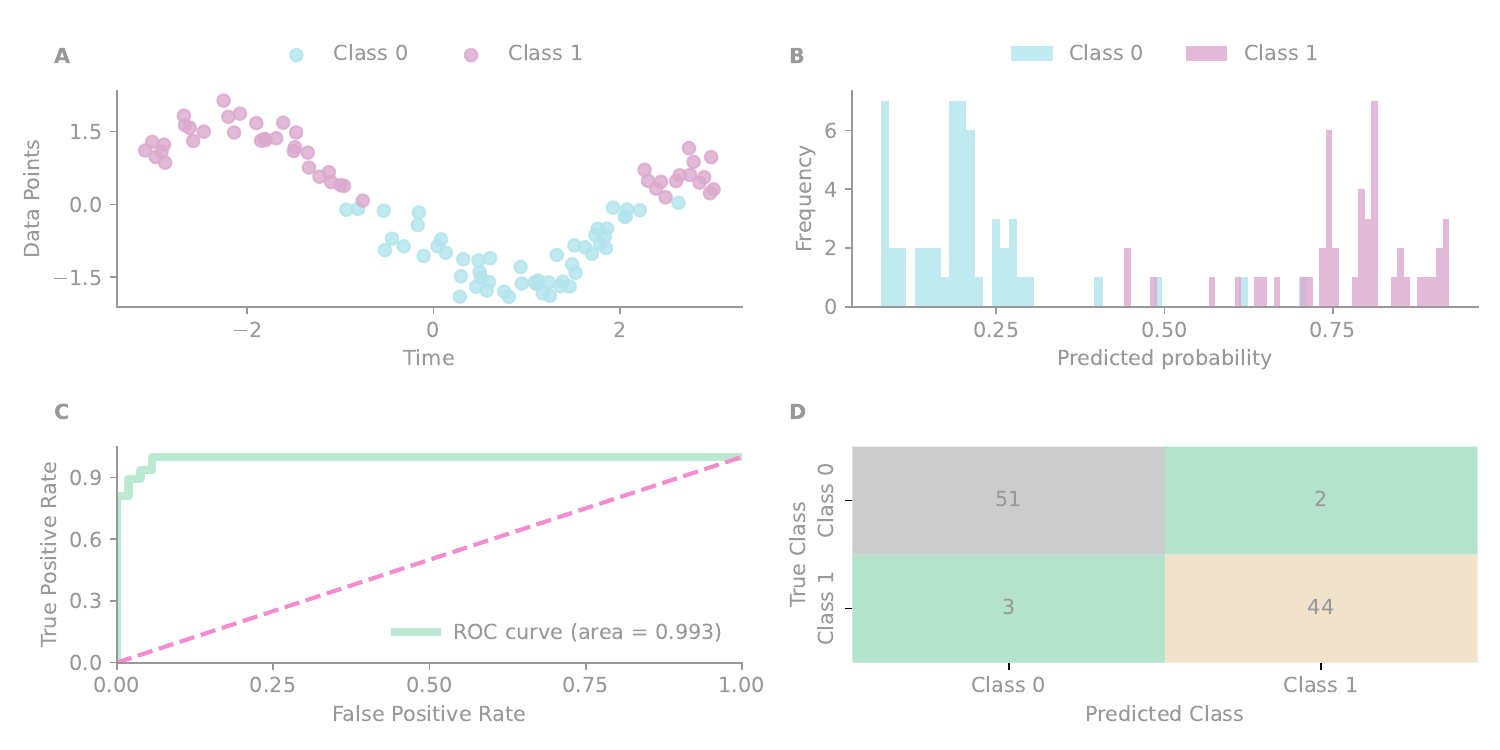}
\caption{Binary classification of the canonical variable $\xi$ from Figure \ref{fig:regression}, with $30\%$ added noise. Classes 0 and 1 correspond to values of $\xi$ below and above the median, respectively. Panel A shows the noisy data points and their classes. Panel B presents the predicted probabilities for both classes after applying a sigmoid function. Panel C depicts the ROC curve with an area under the curve (AUC) of 0.993, indicating excellent model performance over the testing set. Panel D displays the confusion matrix for predictions using a threshold $\Pi_\text{ref} = 1/2$, summarising the classification performance on the testing set.}
\label{fig:classification}
\end{figure*}

Classification can be conceptually understood as a specialised form of regression that discretises the target variable, $\hat{\Xi}$, into $N$ distinct classes, $\{\hat{\Xi}^1, \ldots, \hat{\Xi}^N\}$, each representing a specific aspect of $\hat{\Xi}$ according to arbitrary cutoffs.
The dataset used for regression $\{t,\hat{\Xi}\}$ is then transformed into $\{t, \{\hat{\Xi}^1, \ldots, \hat{\Xi}^N\}\}$.
For instance, in a binary classification scenario, $N=2$, the target variable $\Xi$ is split into two complementary classes: define $\hat{\Xi}^+$ when $\hat{\Xi} \geq 0$, and $\hat{\Xi}^- = 1 - \hat{\Xi}^+$ otherwise.
Therefore, some time points will be labelled as $\hat{\Xi}^+$ and others as $\hat{\Xi}^-$, indicating whether the canonical observable is positive or negative with respect to the zero point reference at those times.
In contexts such as particle motion in an ion trap, an algorithm of this kind would enable measurements to be predicted and grouped based on predefined, discrete classes.

The focus of this discussion will be on binary classification, yet the foundational principles and techniques described here can be adapted to tackle problems involving multiple classes.
In general, to discretise the target variable into $N$ classes, techniques such as one-vs-rest (OvR) or one-vs-one (OvO) can be employed \cite{rifkin04}.
In the OvR strategy, $N$ separate binary classifiers are trained, each one distinguishing a single class from the rest.
Specifically, each classifier is responsible for determining the class membership of a data point.
In the OvO approach, a binary classifier is trained for every possible pair of classes, resulting in $N(N-1)/2$ classifiers.
Each classifier in this strategy distinguishes between two classes, and the final class assignment is determined by combining the outcomes of all pairwise classifiers.

The following binary classification algorithm predicts the discrete classes of $\hat{\Xi}$, namely $\hat{\Xi}^1$ and $\hat{\Xi}^0 = 1 - \hat{\Xi}^1$. 
It optimises the parameters of $\theta_\mathcal{P}(t)$ to maximise the likelihood $\Pi$ of accurately classifying data points as $\hat{\Xi}^1$.
Consequently, $1 - \Pi$ corresponds to the likelihood of classifying data points as $\hat{\Xi}^0$.
An optimal fit of $\mathcal{P}$ thus makes the predicted labels $\Xi^1$ and $\Xi^0$ match the actual classes $\hat{\Xi}^1$ and $\hat{\Xi}^0$, respectively.

The likelihood $\Pi$ is quantified by mapping the trained variable $\Xi$ to a probability regime using the logistic sigmoid function $\sigma(\xi) = 1/(1+e^{-\xi})$. 
Labels are assigned as $\Xi^1$ if $\Pi \geq \Pi_\text{thr}$, where $\Pi = \sigma(\Xi)$, and as $\Xi^0$ otherwise. 
The threshold $\Pi_\text{thr}$, typically set at $1/2$ for a balanced dataset, discriminates between the two classes.

A suitable measure for estimating discrepancies between target and predicted labels is the binary cross-entropy cost function.
This cost function is defined as $\mathcal{C}_C  = -\sum_{j=1}^n \left[\hat{\Xi}_j \log(\Pi_j) + (1-\hat{\Xi}_j) \log(1-\Pi_j)\right]$
where $n$ is the number of data points. 
If the ion's trajectory is the target, and if it is split into two classes through a certain cutoff, the summand in the cost function becomes:
\begin{equation*}
\hat{x}(t_j) \log[\sigma( x(t_j) )] + (1-\hat{x}(t_j)) \log[1-\sigma( x(t_j) )],
\end{equation*}
with $x(t_j)$ estimated via $Q = u(t,-\tau)Q_0$.
Using Bayesian optimisation to minimise $\mathcal{C}_C$ iteratively refines $\theta_\mathcal{P}(t)$, enhancing parameter accuracy by penalising deviations from the actual class labels.

A numerical example illustrates the algorithm’s operation in this context. 
The trajectory $\hat{x}$ shown in Panel A of Figure \ref{fig:regression} is divided into two classes based on the median $\mu$, forming a synthetic dataset for classification: Class 1, $\hat{x}^1$, for $\hat{x} \geq \mu$ and Class 0, $\hat{x}^0$, for $\hat{x} < \mu$. 
This cutoff creates a balanced dataset, negating the need for resampling unless one class significantly outnumbered the other. 
Here, $30\%$ Gaussian noise has been added to the original trajectory to simulate real-world conditions.

The dataset $\{t, \{\hat{x}^1,\hat{x}^0\} \}$ is then divided into training and testing subsets with an 80:20 ratio, resulting in 400 training time points and 100 testing time points.
Panel A of Figure \ref{fig:classification} shows the testing subset.
The training set initiates the parameter fitting process for the classification model, using the same ansatz $\theta_\mathcal{P}(t)$ as in regression.

After 52 iterations, the model’s performance on the training subset achieves a receiver operating characteristic (ROC) curve score of 0.99, indicating an almost perfect fit.
The model is then evaluated on the testing subset, achieving an ROC score of $0.993$, demonstrating good generalisation to unseen data.
Figure \ref{fig:classification} summarises these results.

At the physical level, when the parameters in $\theta_\mathcal{P}(t)$ have been optimised, the implemented elastic field $\beta(t)$ permits learning operations at the quantum level.
In this setup, the learning algorithm is implemented in the potential field of the ion trap itself, with predictions driven by the dynamics therein.
Therefore, this scheme represents a form of quantum machine learning where, rather than optimising the parameters of quantum gates, the elastic field allows learning operations to drive observables.

For regression tasks, any ion introduced into the trap will manifest motion consistent with the learned target data, assuming initial conditions are established. 
Additionally, given that the target data may be either empirical or synthetic, the synthetic option facilitates the design of specific $\beta(t)$ profiles to exert precise control over particle dynamics. 
In classification scenarios, comparable benefits are observed. 
Classifying temporal instances when a particle remains within a specified spatial region---or transitions to another state---can result in predictions with reasonable accuracy.

%
%
\begin{acknowledgments}
The author would like to thank the Luxembourg Fonds National de la Recherche for the funding that made this research possible. Special thanks to Prof. J. Gon\c{c}alves for hosting this learner in the Systems Control Group.
\end{acknowledgments}
%
%
\bibliography{bibliography.bib}
%
%
\end{document}